\documentclass[aps,prl,reprint,superscriptaddress]{revtex4-2}
\usepackage{amsmath}
\usepackage{amsfonts}
\usepackage{xcolor,graphicx}        
\usepackage{braket}
\usepackage{float}
\usepackage{tikz}
\usepackage{hyperref}
\usepackage{booktabs}  
\usepackage{multirow}  
\usepackage{amsmath}    
\usepackage{array}
\hypersetup{
	colorlinks = true,
	linkcolor = cyan,
	anchorcolor = cyan,
	citecolor = cyan,
	filecolor = cyan,
	urlcolor = cyan}
\begin{document}
\title{Operator-based propagation of {W}hittaker and {H}elmholtz–{G}auss beams}

\author{M. A. Jácome-Silva}
\affiliation{Instituto Nacional de Astrofísica Óptica y Electrónica (INAOE)\\ Luis Enrique Erro 1, Santa María Tonantzintla, Puebla, 72840, Mexico}

\author{I. Julían-Macías}
\affiliation{Instituto Nacional de Astrofísica Óptica y Electrónica (INAOE)\\ Luis Enrique Erro 1, Santa María Tonantzintla, Puebla, 72840, Mexico}
\email[e-mail:\,]{cavalierjulian86@gmail.com}

\author{F. Soto-Eguibar}
\affiliation{Instituto Nacional de Astrofísica Óptica y Electrónica (INAOE)\\ Luis Enrique Erro 1, Santa María Tonantzintla, Puebla, 72840, Mexico}

\author{U. Ruiz-Corona}
\affiliation{Instituto Nacional de Astrofísica Óptica y Electrónica (INAOE)\\ Luis Enrique Erro 1, Santa María Tonantzintla, Puebla, 72840, Mexico}

\author{I. Ramos-Prieto}
\affiliation{Instituto Nacional de Astrofísica Óptica y Electrónica (INAOE)\\ Luis Enrique Erro 1, Santa María Tonantzintla, Puebla, 72840, Mexico}

\author{D. {Sánchez}-{de-la-Llave}}
\affiliation{Instituto Nacional de Astrofísica Óptica y Electrónica (INAOE)\\ Luis Enrique Erro 1, Santa María Tonantzintla, Puebla, 72840, Mexico}

\author{H. M. Moya-Cessa}
\affiliation{Instituto Nacional de Astrofísica Óptica y Electrónica (INAOE)\\ Luis Enrique Erro 1, Santa María Tonantzintla, Puebla, 72840, Mexico}

\begin{abstract}
We introduce a compact operator-based technique that solves the paraxial wave equation for a broad class of structured light fields. Using the spatial evolution operator to propagate two families of physically apodized inputs, Gaussian-apodized Whittaker integrals, and Gaussian-apodized Helmholtz fields, we derive closed-form expressions that retain the Gaussian width and therefore describe finite-energy beams. The method unifies and extends the Helmholtz–Gauss families and readily generalizes to nonseparable nondiffracting architectures. Experiments on superposed Bessel–Gauss beams confirm the predicted transverse rotations, demonstrating that this operator approach is a fast, transparent, and practical alternative to standard diffraction-integral treatments.
\end{abstract}

\maketitle
The discovery of diffraction-free Bessel beams by Durnin et al.~\cite{DurninDiffraction, DurninExact} opened new avenues for understanding light propagation. Subsequently, other nondiffracting families emerged, Mathieu~\cite{RenGeneration} and parabolic beams~\cite{LiuOptical}, each associated with distinct coordinate systems (cylindrical circular, elliptic, and parabolic, respectively). These separable solutions carry orbital angular momentum transferable to particles~\cite{BouchalNondiffracting, LopezOrbital, LiuOptical}, enabling optical manipulation applications. Despite originating from different coordinate systems, all can be unified through the Whittaker integral~\cite{WhittakerBook}, which describes nondiffracting beams as plane-wave superpositions weighted by an angular spectrum. This framework has recently allowed the creation of complex nonseparable beam structures~\cite{CabreraDurnin, AmaralMethod, ZannottiShaping, LanCustomizing, HuOrbital}.

Under paraxial propagation, nondiffracting beams acquire a Gaussian envelope, ensuring finite energy. Gori et al.~\cite{GoriBesselGauss} introduced the first such beam, the Bessel–Gauss, by multiplying a Bessel function with a Gaussian profile. Extensions to Mathieu–Gauss and parabolic–Gauss beams~\cite{GutierrezHelmholtz, HernandezExperimental, NguyenOn-demand} followed, preserving their nonparaxial geometries while becoming experimentally accessible via spatial light modulators~\cite{LopezObservation2}. Chen et al.~\cite{ChenGeneration} further applied Fresnel diffraction and stationary phase methods to generate paraxial versions of Zannotti's structured caustic beams~\cite{ZannottiBook} using phase plates and axicons.

The operator-technique method has recently been proven to be effective in solving the paraxial wave equation with various initial field distributions~\cite{MoyaCauchy, KorneevAsymmetric, MoyaParaxial}. Here, we apply this approach to two Gaussian-apodized initial conditions at $z=0$: (i) a Gaussian-modulated Whittaker integral and (ii) a Gaussian-modulated Helmholtz solution. We demonstrate that both formulations yield consistent results and provide an alternative derivation of the Helmholtz-Gauss beam family (Bessel-Gauss, Mathieu-Gauss, and parabolic-Gauss).

\textit{Whittaker-Gauss beams.}
In the paraxial approximation, the free-space light propagation obeys
\begin{equation}
    \nabla_\perp^2 \psi(\mathbf{r}_\perp,\tau) + 2i\frac{\partial \psi(\mathbf{r}_\perp,\tau)}{\partial \tau} = 0,
    \label{paraxialequation}
\end{equation}
where $\tau = z/k$ is the reduced axial coordinate, $\mathbf{r}_\perp = (x,y)$ the transverse position, $\nabla_\perp^2$ the transverse Laplacian, and $k = 2\pi/\lambda$ the wavenumber. Exploiting the structural analogy with the Schrödinger equation, the solution admits the operator representation~\cite{Stoler}: $\psi(\mathbf{r}_\perp,\tau) = \exp\left(\tfrac{i\tau}{2} \nabla_\perp^2\right) \psi(\mathbf{r}_\perp,0)$, where the exponential acts as the spatial evolution operator in the initial field $\psi(\mathbf{r}_\perp,0)$. We consider a Gaussian-apodized Whittaker integral:
\begin{equation}
    \psi(\mathbf{r}_\perp,0) = e^{-g(x^2 + y^2)} \int\limits_{-\pi}^{\pi} A(\varphi) e^{i\mathbf{k}_\perp \cdot\, \mathbf{r}_\perp} d\varphi,
    \label{initialfieldW}
\end{equation}
with real width parameter $g$ and $\mathbf{k}_\perp = k_\perp(\cos\varphi\,\hat{\mathbf{x}} + \sin\varphi\,\hat{\mathbf{y}})$. The integral represents a coherent superposition of plane waves with wavevectors in a circle, weighted by the angular spectrum $A(\varphi)$, while the Gaussian factor ensures finite energy.

Applying $\exp\left(\tfrac{i\tau}{2} \nabla_\perp^2\right)$ to \eqref{initialfieldW} and using the canonical commutation relations $[\partial_x, x] = [\partial_y, y] = 1$, we perform a Lie-algebraic similarity transformation~\cite{MoyaCauchy, Wei1964} on the Gaussian envelope, yielding
\begin{equation}
\label{Gaussian}
\begin{split}
e^{\frac{i\tau}{2} \nabla_\perp^2} & e^{-g(x^2 + y^2)} e^{-\tfrac{i\tau}{2} \nabla_\perp^2} = e^{\alpha(\tau)(x^2 + y^2)} \\
&\times e^{\beta(\tau) \left( x\tfrac{\partial}{\partial x} + \tfrac{\partial}{\partial x}x + y\tfrac{\partial}{\partial y} + \tfrac{\partial}{\partial y}y \right)} e^{\gamma(\tau)\nabla_\perp^2},
\end{split}
\end{equation}
with $\alpha(\tau) = -g/\omega(\tau)$, $\gamma(\tau) = -g\tau/\omega(\tau)$, $\beta(\tau) = -\pi/4 - i\ln(i\omega(\tau))/2$, and $\omega(\tau) = 1 + 2ig\tau$. The term $\beta(\tau)$ is the squeeze operator that induces coordinate rescaling. Crucially, the plane-wave components in \eqref{initialfieldW} are eigenfunctions of the transverse Laplacian with eigenvalue $-k_\perp^2$:
\begin{equation}
    \nabla_\perp^2 e^{i\mathbf{k}_\perp \cdot\,\mathbf{r}_\perp} = -k_\perp^2 e^{i\mathbf{k}_\perp \cdot\,\mathbf{r}_\perp}.
    \label{eigenfunction}
\end{equation}
This property allows the operator $\exp[\gamma(\tau)\nabla_\perp^2]$ in \eqref{Gaussian} to act simply as a multiplicative phase factor $\exp[-\gamma(\tau)k_\perp^2]$ on each plane wave. Combining the factorized operators and applying the squeeze transformation yields the propagated field:
\begin{equation}
\label{WhiatterGaussbeam}
\begin{split}
    \psi(\mathbf{r}_\perp,\tau) = & \frac{e^{-k_\perp^2 \left( \tfrac{g\tau^2}{\omega(\tau)} + \tfrac{i\tau}{2} \right)} e^{-\tfrac{g (x^2 + y^2)}{\omega(\tau)}}}{\omega(\tau)}\\
    & \times \int_{-\pi}^{\pi} A(\varphi) e^{\tfrac{i \mathbf{k}_\perp \cdot \mathbf{r}_\perp}{\omega(\tau)}} d\varphi.
\end{split}
\end{equation}
This expression reveals the dual diffraction mechanism: the Gaussian envelope controls beam expansion and wavefront curvature through $\omega(\tau)$, while the angular spectrum undergoes coordinate renormalization $\mathbf{r}_\perp \to \mathbf{r}_\perp/\omega(\tau)$. Although equivalent to Ref.~\cite{GutierrezHelmholtz}, our operator formalism is more general and naturally incorporates the Gaussian factor, avoiding the nonphysical infinite-energy limit ($g=0$) common in prior treatments~\cite{CabreraDurnin, AmaralMethod, ZannottiShaping, LanCustomizing, HuOrbital}.

Table \ref{table1} lists the functions $A(\varphi)$ that yield separable Helmholtz solutions in Cartesian, polar, elliptic, and parabolic coordinates. Substituting these into \eqref{WhiatterGaussbeam} recovers the Bessel-Gauss, Mathieu-Gauss, and parabolic-Gauss beams. Table \ref{table2} presents $A(\varphi)$ for nondiffracting nonseparable beams—Bessel-lattice, Durnin-Whitney, Archimedes and elliptic structures—obtained by prescribing specific caustic geometries or wavevector curves.

\begin{table}[ht]
\centering
\caption{Expression of $A(\varphi)$ to generate separable solutions to the Helmholtz equation.}
\label{table1}
\renewcommand{\arraystretch}{1.5} 
\begin{tabular}{ll}
\toprule
\textbf{Beam Type} & \boldmath{$A(\varphi)$} \\
\midrule
Plane Waves & $\sum_{m} \delta(\varphi - \varphi_m)$ \\
Bessel Beam & $e^{im(\varphi - \pi/2)}$ \\
Mathieu Beam & $ce_m(\varphi; q) + i se_m(\varphi; q)$ \\
\midrule
\multirow{4}{*}{Parabolic Beam} & $A_e(\varphi; a) = \frac{1}{2\sqrt{\pi|\sin\varphi|}} e^{ia\ln|\tan(\varphi/2)|}$ \\
\cmidrule{2-2}
 & $A_o(\varphi; a) = \frac{1}{i} \begin{cases} 
-A_e(\varphi; a), & \varphi \in (\pi, 2\pi) \\ 
A_e(\varphi; a), & \varphi \in (0, \pi) 
\end{cases}$ \\
\bottomrule
\end{tabular}
\end{table}

\begin{table}[ht]
\centering
\caption{Expression of $A(\varphi)$ to generate non-diffracting beams as non-separable solutions.}
\label{table2}
\renewcommand{\arraystretch}{1.8} 
\begin{tabular}{ll}
\toprule
\textbf{Beam Type} & \boldmath{$A(\varphi)$} \\
\midrule
Bessel-lattice & $e^{i \left[ m\varphi - \frac{b}{2} \sin(2\varphi) \right]}$ \\
Durnin-Whitney & $e^{i (m\varphi + b\varphi^2)}$ \\
Simple plane curves & $e^{im \tan^{-1} \left[ \frac{k_y(\varphi)}{k_x(\varphi)} \right]}$ \\
Non-diffracting structures & $e^{i a k_\perp \frac{(n-1)^2}{n+1} \cos \left( \frac{n+1}{n-1} \varphi \right)}$ \\
Archimedes structure & $e^{i k_\perp \left[ \frac{b\varphi(\varphi-\pi)}{2} + a\varphi \right]}$ \\
Elliptic structure & $e^{i k_\perp \left[ (a+b)\varphi - \frac{(a+b)}{2} \sin\varphi \right]}$ \\
\bottomrule
\end{tabular}
\end{table}

To demonstrate the analytical results, we generated several fields experimentally. The experimental setup consists of a 4-f optical system, where a linearly polarized He-Ne laser ($\lambda = 632.8 \; $ nm) impinges on a spatial light modulator (SLM). A synthetic phase hologram (SPH)~\cite{Arrizon:07} is displayed on the SLM. A lens ($f_1=40$ cm) performs the SPH Fourier transforms, allowing a spatial filter in the Fourier plane to transmit the desired field. Afterward,  a second lens ($f_2=40$ cm) performs the inverse Fourier transform to recover the encoded field; finally, a CCD records the intensity distribution. In Figs.~\ref{figura1}--\ref{figura3}, we present the analytical and experimental intensity distributions at several representative planes along the propagation direction $z$ for the superposition of two Whittaker-Gauss beams, each characterized by its own angular spectrum given by $A_j(\varphi) = (1/2)\exp[(i(m_j\varphi - \pi/2)]$ and a fixed value ${k_\perp}_j$ ($j = 1,\,2$). We emphasize that substituting this angular expression into the integral of \eqref{WhiatterGaussbeam} yields the factor $J_m(k_\perp\rho/\omega(\tau))e^{im\theta}$ with $\rho=\sqrt{x^{2}+y^{2}}$ and $\theta=\arctan(y/x) $(see Table~\ref{table1}), showing that  the resulting optical field can be expressed as the product of a Bessel function rescaled by a factor $\omega(\tau)$ and a phase.

As a first instance, Figs. \ref{figura1}($\mathrm{a}_1$)-($\mathrm{c}_1$) show the analytical intensity distribution of the superposition of Bessel-Gauss beams, at three transverse planes: ($\mathrm{a}_1$) in $z = 0 \,\mathrm{m}$, ($\mathrm{b}_1$) in $z = 0.25 \, \mathrm{m}$ and ($\mathrm{c}_1$) in $z = 0.75 \, \mathrm{m}$,  where $m_1 = 1$, ${k_\perp}_1 = 1800 \, \mathrm{m}^{-1}$, $m_2 = -1$ and ${k_\perp}_2 = 8900 \, \mathrm{m}^{-1}$. Figs.~\ref{figura1}($\mathrm{a}_2$)-($\mathrm{c}_2$) correspond to the experimental intensity distributions in $z \approx 0 \, \mathrm{m}$, $z \approx .25 \, \mathrm{m}$ and $z \approx .75 \, \mathrm{m}$. From these graphs, a relationship can be observed between the values of $m_1$, $m_2$, and the number of central lobes, along with a rotation in the $(x,y)$ plane whose direction depends on the transverse wavenumber $k_\perp$. Additionally, if the values of $k_\perp$ are equal, this rotation does not occur, although two central lobes are still generated. 
\begin{figure}[h!]
\centering\includegraphics[width=\linewidth]{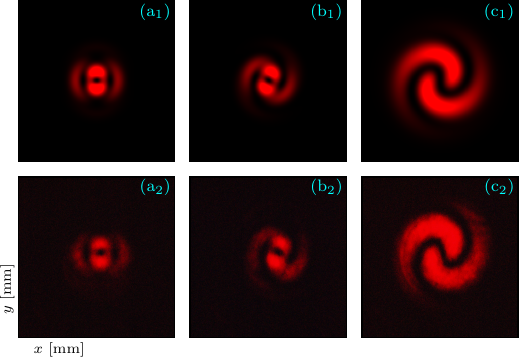}
\caption{Intensity distributions for the superposition of Bessel-Gauss beams which are characterized by $(m_1 = 1, \, {k_\perp}_1 = 1800 \, \mathrm{m}^{-1})$, and $(m_2 = -1, \, {k_\perp}_2 = 8900 \, \mathrm{m}^{-1})$, at three transverse planes: ($\mathrm{a}_1$) at $z = 0.00 \,  \mathrm{m}$, ($\mathrm{b}_1$) at $z = 0.25 \,  \mathrm{m}$ and ($\mathrm{c}_1$) at $z = 0.75 \, \mathrm{m}$. The corresponding experimental distributions are shown in ($\mathrm{a}_2$)-($\mathrm{c}_2$). The experimental parameters are $g = 0.25 \times 10^{7} \,  \mathrm{m}^{-2}$ and $\lambda = 632.8 \, \mathrm{nm}$, all within a viewing window of $4 \, \mathrm{mm}$.}
\label{figura1}
\end{figure}
As a second instance, in the first row of Fig.~\ref{figura2} we show the analytical intensity distribution of the superposition of Bessel-Gauss beams in three transverse planes: ($\mathrm{a}_1$) in $z = 0.00 \, \mathrm{m}$, ($\mathrm{b}_1$) in $z = 0.25 \, \mathrm{m}$ and ($\mathrm{c}_1$) in $z = 0.75 \, \mathrm{m}$, when $m_1 = 1$, ${k_\perp}_1 = 1800 \, \mathrm{m}^{-1}$, $m_2 = 0$ and ${k_\perp}_2 = 8900 \, \mathrm{m}^{-1}$. In the second row of Fig.~\ref{figura2}, the corresponding experimental distribution is presented. In this case, it can be noted that, unlike the symmetric concentric-ring pattern of a single Bessel beam, the superposition of these Bessel functions of different orders creates regions of zero intensity. Again, a rotation in the $(x,y)$ plane is observed, whose direction depends on the transverse wavenumber $k_\perp$. 
\begin{figure}[h!]
\centering\includegraphics[width=\linewidth]{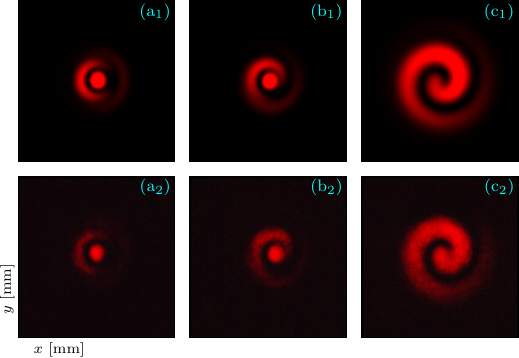}
\caption{Intensity distributions for the superposition of Bessel-Gauss beams which are characterized by $(m_1 = 1, \, {k_\perp}_1 = 1800 \, \mathrm{m}^{-1})$, and $(m_2 = 0, \, {k_\perp}_2 = 8900 \,  \mathrm{m}^{-1})$, at the same planes as in Fig.~\ref{figura1}. The corresponding experimental distributions are shown in ($\mathrm{a}_2$)-($\mathrm{c}_2$). The parameters of $g$ and $\lambda$, as well as the observation window width are the same as those in the Fig.~\ref{figura1}.}
\label{figura2}
\end{figure}
\begin{figure}[h!]
\centering\includegraphics[width=\linewidth]{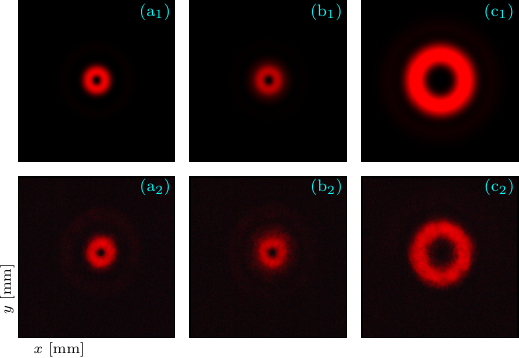}
\caption{Intensity distributions for the superposition of Bessel-Gauss beams which are determined by $(m_1 = 1, \, {k_\perp}_1 = 1800 \, \mathrm{m}^{-1})$, and $(m_2 = 1, \, {k_\perp}_2 = 8900 \, \mathrm{m}^{-1})$, at the same planes as in Fig.~\ref{figura1}. The corresponding experimental distributions are shown in ($\mathrm{a}_2$)-($\mathrm{c}_2$). The parameters of $g$ and $\lambda$, as well as the observation window width are the same as those in the Fig.~\ref{figura1}.}
\label{figura3}
\end{figure}
Finally, in Figs.~\ref{figura3}($\mathrm{a}_1$)-($\mathrm{c}_1$), the analytical distribution of the superposition of Bessel-Gauss beams can be observed in three transverse planes: ($\mathrm{a}_1$) in $z = 0.00 \, \mathrm{m}$, ($\mathrm{b}_1$) in $z = 0.25 \, \mathrm{m}$, and ($\mathrm{c}_1$) in $z = 0.75 \, \mathrm{m}$, for $m_1 = 1$, ${k_\perp}_1 = 1800 \, \mathrm{m}^{-1}$, $m_2 = 1$ and ${k_\perp}_2 = 8900 \, \mathrm{m}^{-1}$. The experimental distributions are shown in Figs.~\ref{figura3}($\mathrm{a}_2$)-($\mathrm{c}_2$). From this last case, we note that the superposition of these beams exhibits a symmetric concentric pattern, and we can observe in all three cases that the intensity distributions diffract as they propagate through different distances $z$. This diffraction occurs due to the Gaussian modulation, which spatially confines the field—clearly differentiating it from its non-diffracting counterpart—and makes it possible to generate these optical fields through phase holography. Furthermore, it does not matter whether the Whittaker integral or Bessel functions are used to generate the phase holograms in the experimental setup or in the numerical simulations, the intensity pattern obtained is the same in both cases. As the three figures illustrate, the close correspondence between theoretical predictions and experimental results validates both the operator-based approach used to derive the theoretical model and the efficacy of the optical field synthesis technique implemented via phase holography. 

\textit{Helmholtz-Gauss beams.} As established, the general expression in \eqref{WhiatterGaussbeam} encompasses the well-known Bessel-Gauss, Mathieu-Gauss, and Parabolic-Gauss beams~\cite{GutierrezHelmholtz}, which are retrieved by prescribing the specific angular spectra listed in Table \ref{table1}. In addition to this integral representation, we now introduce a direct derivation utilizing the operator formalism~\cite{Stoler, MoyaCauchy}. For this purpose, we define the initial amplitude as a Gaussian-apodized Helmholtz field:
\begin{equation}
   \psi(\mathbf{r}_\perp,0) = e^{-g(x^2 + y^2)}\Phi(x,y),\label{initialfieldH}
\end{equation}

where the function $\Phi(x,y)$ satisfies the transverse Helmholtz equation $\nabla_\perp^2\Phi = -k_\perp^2 \Phi$. To determine the propagation dynamics, we apply the evolution operator to this initial profile. Since the Gaussian factor is structurally identical to the one analyzed in the previous section, we can bypass the detailed operator disentanglement. By invoking the same similarity transformation \eqref{Gaussian}, derived from the Hadamard lemma and the Wei-Norman theorem~\cite{Wei1964}, the evolution is governed by the action of the factorized operators on the Helmholtz function $\Phi(x,y)$.

Two mechanisms dictate the final form of the field: first, the operator $\exp[\gamma(\tau) \nabla_\perp^2]$ acts on $\Phi(x,y)$ simply as a multiplicative phase factor $\exp[-\gamma(\tau)k_\perp^2]$, given that $\Phi$ is an eigenfunction of the transverse Laplacian operator ($\nabla_\perp^2\Phi = -k_\perp^2 \Phi$). Second, the remaining operator term, identified as the squeeze operator $\exp \left[\beta(\tau)(q\partial_q + \partial_q q)\right]$ (with $q=x,y$), induces global coordinate renormalization. This operation scales the function arguments to $q \to q/\omega(\tau)$ and introduces an overall amplitude factor $1/\omega(\tau)$. Combining these contributions leads directly to the closed-form solution:
\begin{equation}\label{propagatedfieldH3}
    \psi(\mathbf{r}_\perp,\tau) = \frac{e^{-k_\perp^2  \left( \tfrac{g\tau^2}{\omega(\tau)} + \tfrac{i\tau}{2} \right)} e^{-\tfrac{g( x^2 + y^2 )}{\omega(\tau)}}}{\omega(\tau)}   \Phi\left(\tfrac{x}{\omega(\tau)},\tfrac{y}{\omega(\tau)}\right).
\end{equation}
To validate this result and demonstrate its equivalence with standard Helmholtz-Gauss beams~\cite{GutierrezHelmholtz}, we consider specific solutions $\Phi(x, y)$ to the 2D Helmholtz equation. We recall that this equation admits the separation of variables in various coordinate systems, giving rise to distinct families of structured light fields. For example, in polar coordinates $(x, y) = (\rho \cos\theta, \rho\sin\theta)$, a fundamental solution is $\Phi(\rho, \theta) = J_m(k_\perp \rho)e^{im\theta}$, where $J_m$ is the Bessel function of the first kind of order $m$. Inserting this ansatz into \eqref{propagatedfieldH3} immediately recovers the well-known Bessel-Gauss beam solution. Furthermore, with respect to Cartesian representation, the general solution can be expressed via the reduced Whittaker integral $\Phi(x, y) = \int_{-\pi}^{\pi} A(\varphi) \, \exp\left[i k_\perp (x \cos\varphi + y \sin\varphi)\right] d\varphi$. By substituting this plane-wave expansion into \eqref{propagatedfieldH3} and exploiting the linearity of the operators, we recover the Whittaker-Gauss beam expression derived in the previous section, thus confirming the consistency of the operator approach. It is well known that the two-dimensional Helmholtz equation admits separable solutions in polar, elliptic, and parabolic coordinates. For example, in polar coordinates, the solution of the Helmholtz equation is given by $\Phi(\rho,\theta) = J_n(k_\perp \rho) e^{i n \theta},$ where $x = \rho \cos\theta$ and $y = \rho \sin\theta$, so substituting this solution into \eqref{propagatedfieldH3} results in the Bessel-Gauss beam. Similarly, by applying the corresponding coordinate transformations, the Mathieu-Gauss and Parabolic-Gauss transformations are obtained.

\textit{Conclusions.} 
We have derived the solutions to the paraxial wave equation, considering two Gaussian-apodized initial amplitudes at $z = 0$, applying an operator-technique from quantum optics. The first initial condition is given by a Gaussian-modulated Whittaker integral, while the second corresponds to a Gaussian-modulated Helmholtz solution, which can be physically realizable due to the Gaussian factor. A key advantage of this operator-based approach is its direct applicability, which is less complex than traditional methods. Furthermore, the methodology determines the solution naturally without requiring an assumption of its specific form, which is both a conceptual and a practical advantage. The first solution obtained in this work, the Gaussian-modulated Whittaker integral, is consistent with that obtained in \cite{GutierrezHelmholtz}, which has contained separable solutions known as Bessel-Gauss, Mathieu-Gauss, and Parabolic-Gauss beams, which are particular cases, since there are a large number of beams determined by their angular spectrum. Although it does not always admit an analytical solution, it can be solved numerically using techniques such as Riemann sums or more advanced quadrature methods. The second solution obtained in this work, the Gaussian-modulated Helmholtz solution, is an alternative form to obtain the separable solutions mentioned above, showing that the operator-technique is a powerful tool. In addition to the above, we presented analytical and experimental results of the intensity distribution, at three different values of z, of the superposition of two Bessel-Gauss beams determined by $A_j(\varphi) = (1/2)\exp[(i(m_j\varphi - \pi/2)]$ and a fixed value ${k_\perp}_j$ ($j = 1,\,2$), showing particular characteristics. By selecting appropriate values, it is possible to create beams that rotate in the transverse $(x, y)$ plane—a phenomenon that has significant implications in modern optics for applications such as particle transport, trapping and manipulation~\cite{BiomedicalAplica, article}. These beams are also used in medical imaging and microscopy~\cite{article, BiomedicalAplica,AplicBessel}, as well as in atmospheric optical and quantum communications~\cite{article}. This opens a wide variety of practical applications for future research where precise control over the wavefront structure is essential.

\date{\today}
\maketitle
\end{document}